# Construction and Observation of Flexibly Controllable High-Dimensional Non-Hermitian Skin Effects

Qicheng Zhang,[†] Yufei Leng,[†] Liwei Xiong, Yuzeng Li, Kun Zhang, Liangjun Qi, and Chunyin Qiu[*]

Key Laboratory of Artificial Micro- and Nano-Structures of Ministry of Education
and School of Physics and Technology, Wuhan University, Wuhan 430072, China
[†]These authors contributed equally: Qicheng Zhang, Yufei Leng
[*]To whom correspondence should be addressed: cyqiu@whu.edu.cn

Non-Hermitian skin effect (NHSE) is one of the most fundamental phenomena in non-Hermitian physics. Although it is established that one-dimensional NHSE originates from the nontrivial spectral winding topology, the topological origin behind the higher-dimensional NHSE remains unclear so far. This poses a substantial challenge in constructing and manipulating high-dimensional NHSEs. Here, an intuitive bottom-to-top scheme to construct high-dimensional NHSEs is proposed, through assembling multiple independent one-dimensional NHSEs. Not only the elusive high-dimensional NHSEs can be effectively predicted from the well-defined one-dimensional spectral winding topologies, but also the high-dimensional generalized Brillouin zones can be directly synthesized from the one-dimensional counterparts. As examples, two two-dimensional nonreciprocal acoustic metamaterials are experimentally implemented to demonstrate highly controllable multi-polar NHSEs and hybrid skin-topological effects, where the sound fields can be frequency-selectively localized at any desired corners and boundaries. These results offer a practicable strategy for engineering high-dimensional NHSEs, which could boost advanced applications such as selective filters and directional amplifiers.

## 1. Instruction

Non-Hermiticity enriches topological physics beyond the existing Hermitian framework [1-6]. One of the most prototypical examples is the non-Hermitian skin effect (NHSE), which refers to the accumulation of an extensive number of bulk eigenstates toward open boundaries [7-24]. For one-dimensional (1D) systems, it is well-established that the NHSE under open boundary conditions (OBCs) originates from the nontrivial point-gap or spectral winding topologies under periodic boundary conditions (PBCs) [16-18]. To date, 1D NHSEs have been extensively demonstrated in various physical platforms [25-34], which sparks potential applications in sensing, funneling and lasering, etc. [35-37]. However, the topological origin behind higher-dimensional NHSEs remains elusive, due to the vast diversity of spectral morphologies and boundary conditions. Very recently, some pioneering works have shed light on the understanding of NHSEs in higher dimensions. For instance, the criteria of spectral area and dynamical degeneracy splitting have been introduced against occurrence of geometry-dependent NHSEs. A generalized Brillouin zone (GBZ) formulation has been developed to calculate the energy spectrum and eigenstate profiles. Additionally, a uniform spectra theorem has been proposed to reveal the relations among skin states on different lattice geometries. It is noteworthy that these theoretical advancements have primarily aimed to elucidate the physical origins of high-dimensional NHSEs, rather than offering a general and practical route to construct and manipulate the NHSEs of desired morphologies. Meanwhile, the associated experimental investigations have mostly been conducted on specific models and case-by-case scenarios, such as the geometry-dependent skin effects [43-45], hybrid skin-topological effects [46-48], Floquet-engineered skin effects [49,50], and non-Hermitian chiral skin effects [51,52].

Here, we propose an intuitive bottom-to-top scheme for constructing high-dimensional non-Hermitian lattices, and exemplify diverse morphologies inherent in high-dimensional NHSEs. In our approach, the high-dimensional non-Hermitian lattices are built by assembling multiple sets of 1D lattices orthogonally, in which their spectra and eigenstates correspond to the Minkowski sums and Kronecker products of the constituent 1D ones, respectively. To validate the effectiveness of our scheme, we conceive two concrete two-dimensional (2D) lattice models. The first model, protected by the twisted spectral winding topologies in two directions independently, exhibits multi-polar NHSEs where the skin states can frequency-selectively localize at any conners of the finite lattice. By contrast, under the interplay of twisted spectral winding topology and nontrivial Hermitian Bloch band topology, the second model shows hybrid skin-topological states that simultaneously localize toward two adjacent conners. Such skin states of different morphologies greatly enrich the scenarios of the existing NHSEs. Experimentally, we successfully implement two 2D nonreciprocal acoustic metamaterials by integrating the cavity-tube structures with highly-engineered unidirectional couplers. For both the NHSEs of distinct topological origins, the anticipated sound field localizations are precisely captured under desired frequencies. These exceptional properties endow our acoustic metamaterials with potential applications in designing advanced devices such as selective filters and directional amplifiers.

## 2. Results and Discussion

### 2.1. Theoretical scheme

Recently, an effective approach has been proposed for building (Hermitian) higher-order topological phases from the established 1D topological systems [53,54]. Inspired by this advancement, we delve into the construction and manipulation of high-dimensional NHSEs in non-Hermitian systems. For simplicity, we focus on the 2D systems. Considering two finite 1D non-Hermitian lattices in the $x$ and $y$ directions, their Hamiltonians satisfy

$$\mathbf{H}_x|\psi_m^x\rangle = \varepsilon_m^x|\psi_m^x\rangle, \quad (1a)$$

$$\mathbf{H}_y|\psi_n^y\rangle = \varepsilon_n^y|\psi_n^y\rangle, \quad (1b)$$

where $\varepsilon_m^x$ ($\varepsilon_n^y$) and $|\psi_m^x\rangle$ ($|\psi_n^y\rangle$) are the $m$th ($n$th) eigenenergy and eigenstate of $\mathbf{H}_x$ ($\mathbf{H}_y$), respectively. Utilizing the Kronecker product of matrices, a 2D non-Hermitian lattice can be synthesized from the two independent 1D lattices, i.e.

$$\mathbf{H}_{2D} = \mathbf{H}_x \otimes \mathbf{I}_y + \mathbf{I}_x \otimes \mathbf{H}_y, \quad (2)$$

where the identical matrix $\mathbf{I}_x$ ($\mathbf{I}_y$) shares the dimensionality of $\mathbf{H}_y$ ($\mathbf{H}_x$). It satisfies the eigenproblem

$$\mathbf{H}_{2D}(|\psi_m^x\rangle \otimes |\psi_n^y\rangle) = (\varepsilon_m^x + \varepsilon_n^y)(|\psi_m^x\rangle \otimes |\psi_n^y\rangle). \quad (3)$$

That is, the 2D spectrum of $\mathbf{H}_{2D}$ is the Minkowski sum of the two 1D spectra, and the associated eigenstate comes from the direct product of the 1D ones. This enables us to artificially construct a desired 2D skin state $|\psi_m^x\rangle \otimes |\psi_n^y\rangle$, by using two known 1D skin states $|\psi_m^x\rangle$ and $|\psi_n^y\rangle$ as building blocks. Meanwhile, without using the abstruse Amoeba formulation [41], one can directly derive the 2D GBZ that contains the key information of 2D skin states. More specifically, the 2D GBZ can be described by a (conventional) real-valued 2D wavevector $\mathbf{k} = (k_x, k_y)$ with $k_x, k_y \in [-\pi, \pi]$, plus a $\mathbf{k}$-dependent 2D imaginary wavevector $i\boldsymbol{\mu}_{2D}(\mathbf{k})$ that essentially characterizes the spatial localization rate of a skin state. According to Eq. (3), it is straightforward to decompose the imaginary one as

$$\boldsymbol{\mu}_{2D}(\mathbf{k}) = \{\mu_x(k_x), \mu_y(k_y)\}, \quad (4)$$

where $\mu_x(k_x)$ and $\mu_y(k_y)$ represent the imaginary wavevectors for the 1D GBZs of $\mathbf{H}_x$ and $\mathbf{H}_y$, respectively. That is, each component of $\boldsymbol{\mu}_{2D}$ depends only one component of $\mathbf{k}$. The proof of Eq. (4) is provided in Section 1of Supporting Information. Naturally, the above approach can be extended to construct the three-dimensional NHSEs with richer morphologies, such as hinge and surface skin states (see Section 2, Supporting Information).

Before delving into the application of our scheme, two aspects are highlighted. On one hand, unlike previous studies that typically start from specific high-dimensional models [xx], our scheme enables the construction from well-established 1D systems. This affords a level of universality across versatile models and NHSEs, such as multipolar NHSEs, hybrid skin-topological effects and multi-interface NHSEs as discussed further below. On the other hand, in contrast to the usual scenario where only one morphology of high-dimensional NHSEs exists in a fixed model [xx], our scheme allows for different morphologies to occur at different frequencies. This flexibility facilitates the manipulation and switching of various morphologies among corner, boundary and interface localizations.

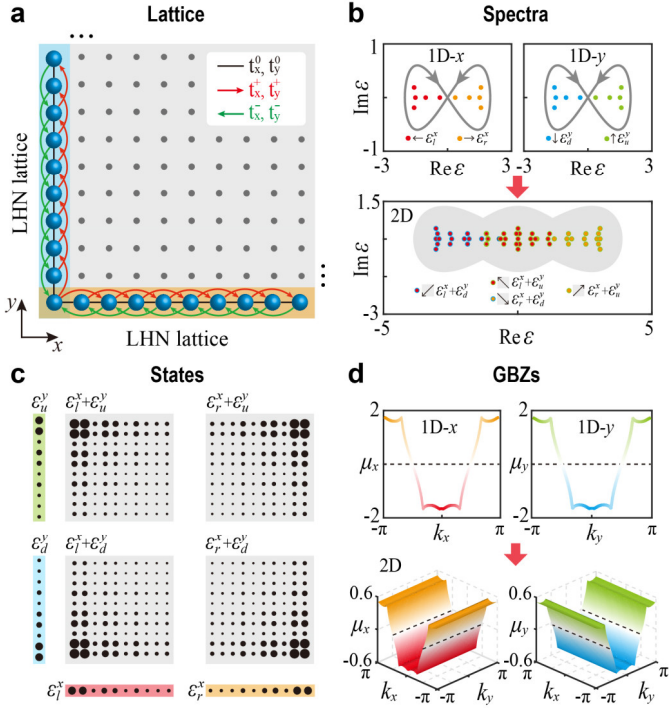

**Figure 1.** 2D model for multi-polar NHSEs. **a** 2D non-Hermitian lattice built from the $x$- and $y$-directed 1D LHN lattices. **b** 2D energy spectra (bottom) generated by summing the 1D ones (top), where thin arrows indicate the localization directions of skin states. The gray lines (area) represent the PBC spectra, while the colored dots represent the OBC spectra for the 1D (2D) systems of 10 ($10 \times 10$) sites. Note that the left-upper and right-lower corner skin states are energetically degenerate in this isotropic construction. **c** Four representative 2D eigenstates formed by the direct products of two sets of 1D eigenstates. The areas of black circles characterize the amplitudes of eigenstates. **d** 2D GBZ (bottom) synthesized from the two 1D ones (top). The hopping parameters used here are $t_x^0 = t_y^0 = 1$ and $t_x^\pm = t_y^\pm = \pm 0.3$.

## 2.2. Constructing 2D multi-polar NHSEs

To demonstrate the effectiveness of our scheme, we first construct a 2D non-Hermitian lattice showcasing highly controllable multi-polar NHSEs. As depicted in Fig. 1a, the lattice is generated by orthogonally compounding two sets of long-range Hatano-Nelson (LHN) lattices [55] along the $x$ and $y$ directions, where $t_x^0$ ($t_y^0$) and $t_x^\pm$ ($t_y^\pm$) represent the reciprocal nearest-neighbor and nonreciprocal long-range hoppings, respectively. We initiate our analysis by examining NHSEs from 1D and 2D complex energy spectra. As shown in Fig. 1b (top), each 1D LHN lattice exhibits a twisted PBC spectrum with two oppositely-oriented loops, and the OBC eigenstates with energies falling inside different loops exhibit opposite skin properties [14]. In general, the skin properties at different eigenenergies of a 2D lattice cannot be simply predicted from its 2D PBC spectrum. However, in our approach, they can be intuitively determined by the two constituent 1D bipolar skin states, as sketched in Fig. 1b (bottom), which exhibit a characteristic of 2D multi-polar corner skin states. Figure 1c provides four typical corner skin states synthesized by the 1D boundary ones. For instance, a 1D leftward skin state at eigenenergy $\varepsilon_l^x$ plus a 1D downward one at $\varepsilon_d^y$ contributes a 2D left-lower corner skin state at $\varepsilon_l^x + \varepsilon_d^y$. Note that the corner skin states at $\varepsilon_l^x + \varepsilon_u^y$ and $\varepsilon_r^x + \varepsilon_d^y$ could be degenerate because of the isotropic construction here, which results in a simultaneous excitation of the left-upper and right-lower corner skin states in practical experiments. Once anisotropy is considered, these corner skin states will be fully frequency-selective because of the broken degeneracy. (In contrast to 2D higher-order topological phases where four topological corner states are usually pinned at the same energy [53,54], here the frequency-selective multi-polar corner skin states enable full controllability on wave localizations.) The skin properties of the synthesized 2D lattice can also be predicted from its 2D GBZ, which is directly obtained from the 1D counterparts $\mu_x(k_x)$ and $\mu_y(k_y)$, as illustrated in Fig. 1d. For each component, the sign of $\mu_x$ or $\mu_y$ corresponds to the direction of the skin state, while the magnitude represents the growth/decay rate of the state. In particular, the sign combinations, $[\text{sgn}(\mu_x), \text{sgn}(\mu_y)] = [-,-], [-,+], [+,-]$ and $[+,+]$ yield the left-lower, left-upper, right-lower and right-upper corner skin states, respectively. Furthermore, the growth/decay direction and rate of each 2D OBC eigenstate are calculated based on the 2D GBZ, as presented in Section 1 of Supporting Information.

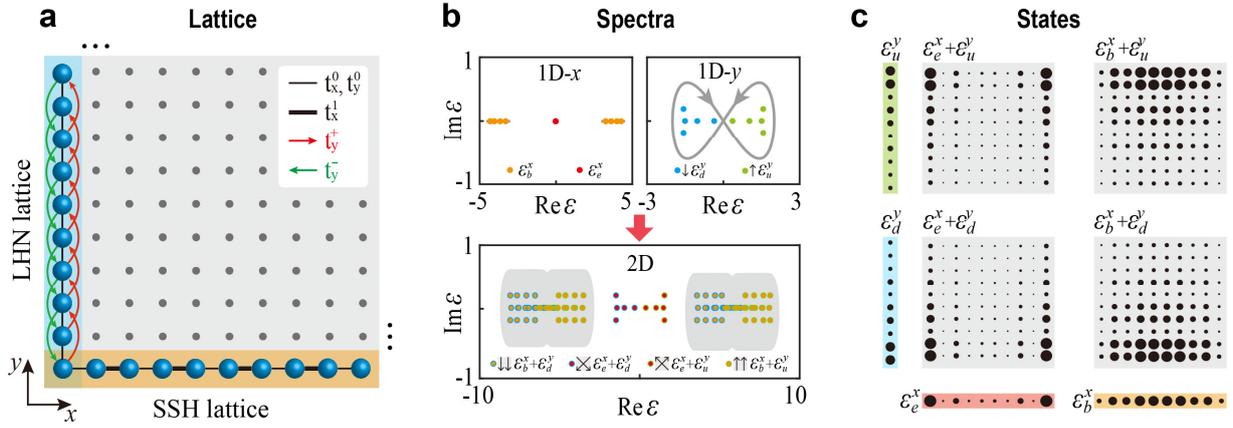

**Figure 2.** 2D model for hybrid skin-topological effects. **a** 2D non-Hermitian lattice built with an *x*-directed SSH lattice and a *y*-directed LHN lattice. **b** 2D energy spectra (bottom) generated by summing the 1D ones (top). The gray lines (areas) represent the PBC spectra, while the colored dots represent the OBC spectra for the 1D (2D) systems of 10 (10 × 10) sites. **c** Four representative 2D eigenstates formed by the direct products of two sets of 1D eigenstates. The hopping parameters used are $t_x^0 = t_y^0 = 1$, $t_x^1 = 6$, and $t_y^{\pm} = \pm 0.3$.

### 2.3. Constructing 2D hybrid skin-topological effects

Our scheme also provides a platform for studying the intriguing interplay of topological states and NHSEs, which greatly enriches high-dimensional robustness beyond (Hermitian) topological phenomena [15]. Here we construct a 2D lattice model to demonstrate hybrid skin-topological states. Comparing with the model in Fig. 1a, now $\mathbf{H}_x$ is replaced by a 1D Hermitian Su-Schrieffer-Heeger (SSH) lattice, which features extended bulk states at $\varepsilon_b^x$ and edge-localized in-gap states at $\varepsilon_e^x$ under OBC. Figures 2b and 2c illustrate the 2D spectra and eigenstates through compounding their 1D counterparts, respectively, manifesting diverse 2D hybrid skin-topological states.

Specifically, the direct product of an *x*-directed topological edge state and a *y*-directed skin state contributes a 2D skin-topological state confined to the two bottom (or top) corners, while the combination of an *x*-directed bulk state and a *y*-directed skin state yields a 2D boundary skin state (Fig. 2c). Notably, unlike the multi-polar corner skin states of number $\mathcal{O}(L^2)$, the hybrid skin-topological corner states are of the number $\mathcal{O}(L)$ [19,20], with $L$ being the length of the lattice. Note that our construction enables a simpler and more intuitive understanding to the hybrid skin-topological effects, compared with those predicted by 2D spectra of different boundary conditions [15,23,24] (see Section 3, Supporting Information).

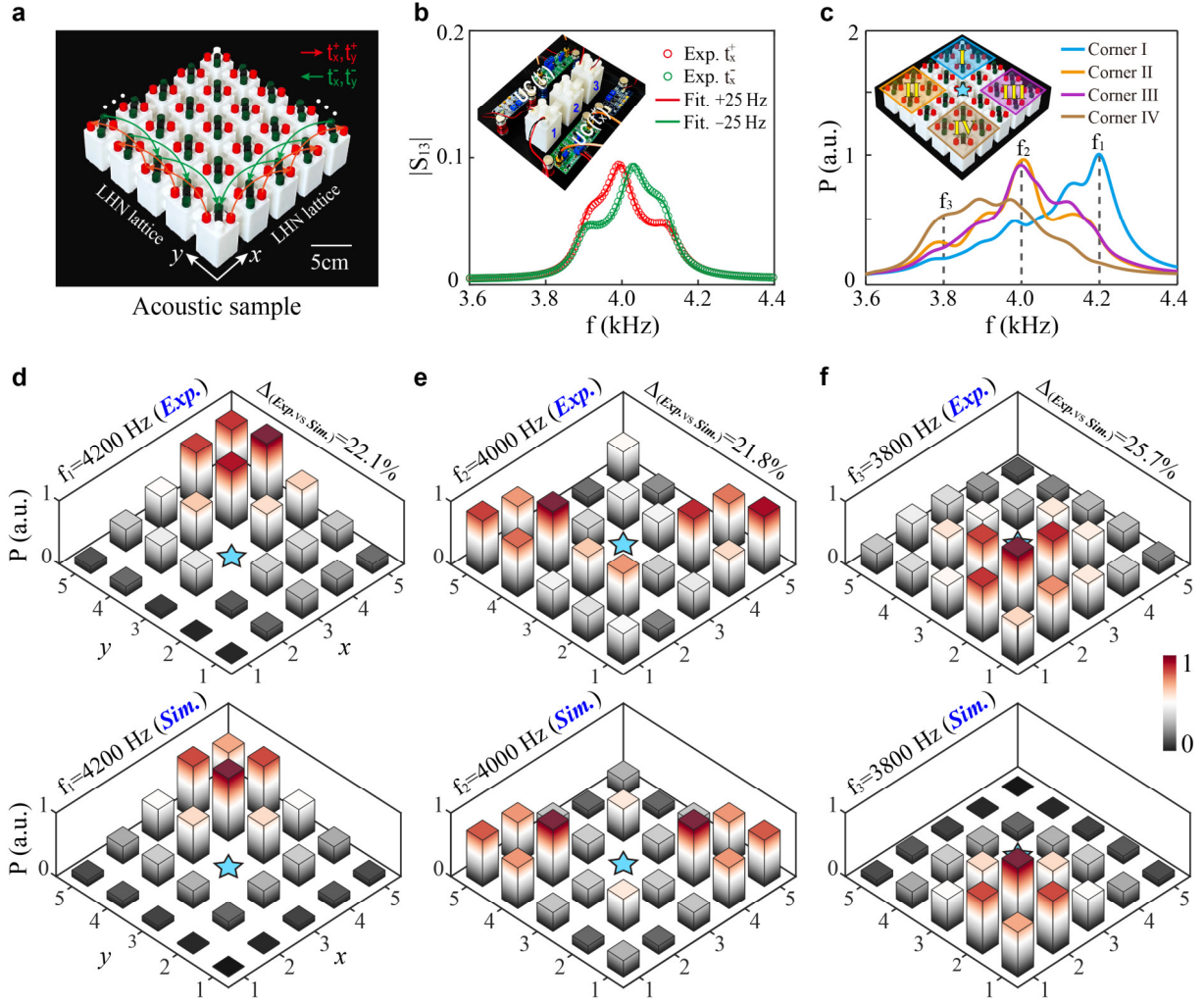

**Figure 3.** Experimental evidence for multi-polar NHSEs. **a** Acoustic metamaterial that emulates the 2D LHN lattice, where the reciprocal nearest-neighbor couplings $t_x^0$ and $t_y^0$ are realized by the passive narrow tubes between the air cavities, and the nonreciprocal long-range couplings $t_x^\pm$ and $t_y^\pm$ are achieved by active unidirectional couplers (Note that only the nonreciprocal couplings on two boundaries are marked with red and green arrows for simplicity. A photograph of the practical sample can be found in Section 4, Supporting Information). **b** Transmission spectra for fitting the nonreciprocal long-range couplings $t_x^\pm$ in a ternary-cavity structure. Inset: photograph of the setup, where UC labels the unidirectional coupler. **c** Sound pressure responses measured for the four sample corners, each averaged by the corresponding four cavities (inset). **d-f** Experimentally measured multi-polar corner skin states at three typical frequencies, together with those simulation results for comparison. In all cases, the sound source (blue star) is positioned in the central cavity of the sample, and the acoustic pressures are normalized by their maximum values individually.

## 2.4. Experiments for multi-polar NHSEs

As shown in Fig. 3a, to experimentally demonstrate the multi-polar NHSEs depicted in Fig. 1, we design and fabricate an acoustic metamaterial of 5 × 5 air-filled cavitives connecting with narrow tubes. The metamaterial can effectively emulates a 2D LHN lattice with complex on-site energy $f_0 \approx (4000 - 50i)$ Hz, reciprocal nearest-neighbor couplings $t_x^0 \approx t_y^0 \approx 78$ Hz, and nonreciprocal long-range couplings $t_x^\pm \approx t_y^\pm \approx \pm 25$ Hz. In particular, the nonreciprocal couplings are realized by active unidirectional couplers [56,57] and their values are determined by fitting the transmission spectra $|S_{13}|$ in a ternary-cavity structure (Fig. 3b). More experimental details are provided in Experimental Section and Supporting Information.

In our acoustic experiments, we place a broadband point source at the central cavity of the metamaterial, and measure the sound pressure responses $P_{(i,j)}(f)$ of all cavities, where $i$ and $j$ refer the cavity indices in the $x$ and $y$ directions, respectively. To better characterize the sound localizations toward different corners, we collect the average pressure of four cavities in each corner region. As shown in Fig. 3c, the pressure responses of different corners dominate within different frequency ranges, experimentally unveiling the characteristics of multi-polar NHSE. Figure 3d presents the sound field distribution at $f_1 = 4200$ Hz. Clearly, the sound field accumalates toward the corner I, aligning with the simulation result obtained through coupled mode theory [see details in Supporting Information and Ref. 58]. A relative deviation $\Delta \approx 22.1\%$, defined by $\Delta = \sqrt{\sum_{i,j}\left[P_{(i,j)}^{Exp} - P_{(i,j)}^{Sim}\right]^2 / \sum_{i,j}\left[P_{(i,j)}^{Sim}\right]^2}$, indicates a good agreement between our experimental and simulation results. Similarly, at $f_3 = 3800$ Hz the sound field accumulates to the corner IV, and at $f_2 = 4000$ Hz, the sound field gathers at the corners II and III simoutanously as expected. These key observations unambiguously evidence the preset 2D multi-polar NHSEs, despite a relatively small sample size used here. (The sound localizations will become more pronounced if employing a larger sample). We also accurately capture the intermediate states between different corner skin states, where three adjacent corners contribute comparably (see Section 5, Supporting Information). Very recently, a pioneering work realizes the manipulation of directional flows underlying 2D NHSEs by varying loss and magnetic field parameters in a quantum walk system [50]. In stark contrast, the frequency selectivity of our 2D skin states enables us to manipulate them without altering any sample parameter.

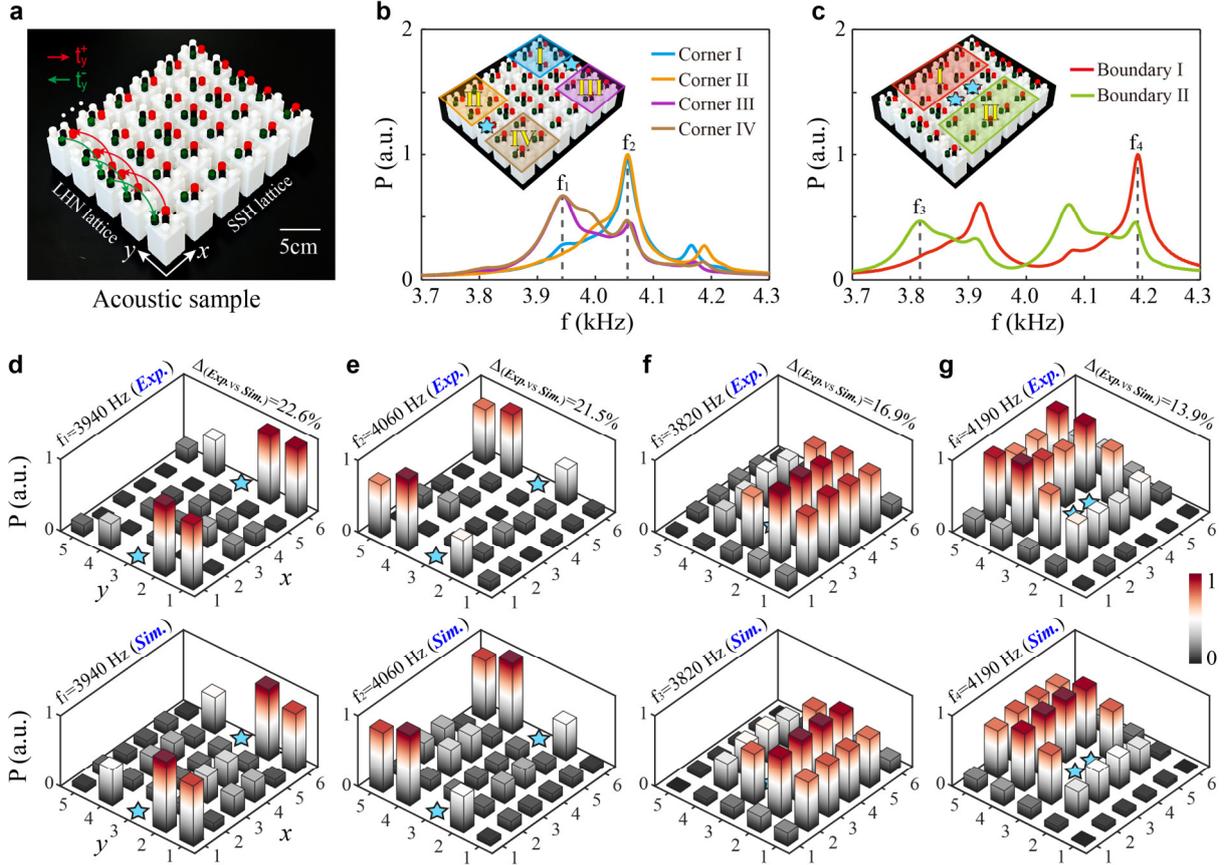

**Figure 4.** Experimental evidence for hybrid skin-topological effects. **a** Acoustic metamaterial that emulates the 2D hybrid lattice. For clarity, only the nonreciprocal couplings $t_y^\pm$ on a $y$-directed boundary are marked with red and green arrows (A photograph of the practical sample is provided in Section 4, Supporting Information). **b** Sound pressure responses measured at four corners of the sample, each of which is averaged by the corresponding four cavities (inset). **c** Sound pressure responses measured at two boundaries, each of which is averaged by the corresponding eight cavities (inset). **d-g** Experimentally measured hybrid skin-topological states at four typical frequencies labeled in **b** and **c**, together with those simulation results for comparison. The blue stars highlight the positions of two sound sources.

### 2.5. Experiments for hybrid skin-topological effects

Next, we experimentally verify the 2D hybrid skin-topological effects conceived in Fig. 2a. The acoustic metamaterial, as shown in Fig. 4a, is comprised of $6 \times 5$ tube-connected air cavities and 30 unidirectional couplers in total. It realizes a 2D hybrid lattice spanned by the $x$-directed SSH lattice and the $y$-directed LHN lattice, where the reciprocal couplings $t_x^0 \approx 40$ Hz, $t_x^1 \approx 140$ Hz, $t_y^0 \approx 45$ Hz, and nonreciprocal couplings $t_y^\pm \approx \pm 25$ Hz.

Experimentally, we place first two identical sources at the central cavities of the two $x$-edges to symmetrically excite the bi-corner skin states hybridized by the $x$-directed topological edge states and the $y$-directed skin states. Figure 4b presents the average spectral responses collected for the four corner regions I-IV. It shows that around the peak frequency $f_1 = 3940$ Hz, the sound pressures at the corners III and IV are much stronger than those of the corners I and II; the situation is reversed near the frequency $f_2 = 4060$ Hz, where the sound responses of the corners I and II become dominant. This spectral characteristic is consistent with Figs. 4d and 4e, the sound fields measured at the peak frequencies $f_1$ and $f_2$, respectively, which exhibit clearly bi-corner skin states localized at different corner pairs. In contrast to those skin-topological states observed previously, which emerge pairwise at diagonal corners [46,47], the 2D hybrid skin-topological states demonstrated here occur at two adjacent corners, accompanying the advantage of frequency-selectivity. We also characterize the 2D boundary skin states resulting from the 1D extended states and skin states in different directions. To do this, we relocate the sound sources to the middle of the sample and collect the average spectral responses for the boundary regions I and II. As expected from Fig. 2b, each spectrum exhibits two prominent peaks at desired frequency windows (Fig. 4c). The sound distributions exemplified at $f_3 = 3820$ Hz (Fig. 4g) and $f_4 = 4190$ Hz (Fig. 4f) further verify the skin effects toward the boundaries I and II, respectively. All experimental results are highly consistent with the simulation ones.

## 3. Conclusion

By extending the 1D NHSEs with well-defined spectral topologies, we have developed a scheme for constructing flexibly controllable high-dimensional NHSEs. The theoretical basis arises from the fact that some high-dimensional non-Hermitian lattices can be mathematically decomposed into multiple independent 1D ones. As concrete examples, we have theoretically proposed and experimentally demonstrated two distinct types of 2D NHSEs: multi-polar corner skin states and hybrid skin-topological states. Such diverse 2D skin states can be selectively excited at different frequencies. In particular, the 2D multi-polar NHSEs, as a long-desired upgrade version of the far-reaching 1D bipolar NHSEs [31], are experimentally observed for the first time. In addition to the high-dimensional NHSEs, our findings also offer efficient routes for demonstrating higher-order band topologies in non-Hermitian systems, such as non-Bloch bulk-boundary correspondence [59] and Floquet second-order topological phases [60,61].

## 4. Experimental Section

*Experimental Samples:* Our experimental samples are 3D-printed by acoustically rigid photosensitive resin with a thickness of 2.0 mm and at a fabrication error of 0.1 mm. Each air-filled cavity has a size of $43.0 \times 28.0 \times 28.0$ mm$^3$. For the sample that emulates the 2D LHN lattice (Fig. 3), each straight connecting tube has a length of 20.3 mm and a cross-sectional area of $7.2 \times 7.2$ mm$^2$. For the sample that realizes the 2D hybrid lattice (Fig. 4), we use thin tubes (of cross-sectional area $5.0 \times 5.0$ mm$^2$ and length 21.0 mm) and thick tubes (of cross-sectional area $9.8 \times 9.8$ mm$^2$ and length 20.5 mm) to mimic the intracell and intercell hoppings of the SSH lattice in the $x$ direction, while use uniform tubes (of cross-sectional area $5.6 \times 5.6$ mm$^2$ and length 20.8 mm) to mimic the LHN lattice in the $y$ direction. To insert the loudspeakers, detectors, and unidirectional couplers, small holes are perforated at the top and bottom sides of the cavities; they are sealed when not in use. The active unidirectional coupler between adjacent cavities comprises a microphone, an amplifier (Type LM386), a phase shifter (Type MCP41010), and a loudspeaker.

*Experimental measurements:* In each measurement, a broadband sound signal (3600∼4400 Hz) is launched from a point-like source driven by the output generator module (B&K Type 3560C). The sound pressure is measured with an acoustic detector (B&K Type 4182) and processed by the signal acquisition module (B&K Type 3560C). For the elementary ternary-cavity experiments in Fig. 3b, we first retrieve the intrinsic parameters $f_0$, $t_x^0$, and $t_y^0$ directly through fitting the transmission spectrum $|S_{13}|$, the sound response of the cavity 1 to the source located at the cavity 3. To achieve desired nonreciprocal couplings, we turn on the active unidirectional couplers and tune the gain-factor/phase-delay of the amplifier/phase-shifter, and retrieve $t_x^\pm$ and $t_y^\pm$ by fitting the transmission spectra $|S_{13}|$ with the known intrinsic parameters.

## Acknowledgements

This project was supported by the National Natural Science Foundation of China (Grants No. 12374418, No. 11890701, and No. 12104346), and the National Key R&D Program of China (Grant No. 2023YFA1406900).

## Conflict of Interest

The authors declare no conflict of interest.

## Keywords

non-Hermitian skin effects, acoustic metamaterials, topological physics, nonreciprocal materials.